# Stellar Companions to TESS Objects of Interest: A Test of Planet–Companion Alignment

Aida Behmard[1] , Fei Dai[1] , and Andrew W. Howard[2]
[1] Division of Geological and Planetary Sciences, California Institute of Technology, Pasadena, CA 91125, USA
[2] Department of Astronomy, California Institute of Technology, Pasadena, CA 91125, USA

## Abstract

We present a catalog of stellar companions to host stars of Transiting Exoplanet Survey Satellite Objects of Interest (TOIs) identified from a marginalized likelihood ratio test that incorporates astrometric data from the Gaia Early Data Release 3 catalog (EDR3). The likelihood ratio is computed using a probabilistic model that incorporates parallax and proper-motion covariances and marginalizes the distances and 3D velocities of stars in order to identify comoving stellar pairs. We find 172 comoving companions to 170 non-false-positive TOI hosts, consisting of 168 systems with two stars and 2 systems with three stars. Among the 170 TOI hosts, 54 harbor confirmed planets that span a wide range of system architectures. We conduct an investigation of the mutual inclinations between the stellar companion and planetary orbits using Gaia EDR3, which is possible because transiting exoplanets must orbit within the line of sight; thus, stellar companion kinematics can constrain mutual inclinations. While the statistical significance of the current sample is weak, we find that $73^{+14}_{-20}\%$ of systems with Kepler-like architectures ($R_P \leqslant 4\ R_\oplus$ and $a < 1$ au) appear to favor a nonisotropic orientation between the planetary and companion orbits with a typical mutual inclination $\alpha$ of $35° \pm 24°$. In contrast, $65^{+20}_{-35}\%$ of systems with close-in giants ($P < 10$ days and $R_P > 4\ R_\oplus$) favor a perpendicular geometry ($\alpha = 89° \pm 21°$) between the planet and companion. Moreover, the close-in giants with large stellar obliquities (planet–host misalignment) are also those that favor significant planet–companion misalignment.

*Unified Astronomy Thesaurus concepts:* Binary stars (154); Exoplanet dynamics (490); Astrometry (80); Bayesian statistics (1900); Hierarchical models (1925)

*Supporting material:* machine-readable tables

## 1. Introduction

Nearly half of FGK stars harbor stellar companions (Raghavan et al. 2010). If exoplanet hosts adhere to this trend, it is likely that many harbor yet-undetected companions whose presence may instigate dynamical processes that drive planet migration, result in misaligned and/or eccentric planetary orbits, or shape system architectures in other ways. Thus, placing observational constraints on the properties of planet host systems with companions is essential for building a complete picture of planet formation and evolution.

Follow-up high-contrast imaging surveys have detected companions to many planet hosts from the Kepler, K2, and Transiting Exoplanet Survey Satellite (TESS) missions (e.g., Dressing et al. 2014; Wang et al. 2015; Kraus et al. 2016; Matson et al. 2018; Ziegler et al. 2020, 2021). More recently, high-precision astrometric measurements from Gaia have made robust identification of comoving stars possible, resulting in large catalogs of stellar companions to planet host stars (e.g., Gaia Collaboration et al. 2018, 2021). For example, Mugrauer (2019), Mugrauer & Michel (2020, 2021), and Michel & Mugrauer (2021) identified companions to confirmed and candidate exoplanet hosts by establishing criteria for proper motions and parallaxes that, when met, constitute high confidence that a pair of stars is gravitationally bound.

Other studies have established probabilistic models to identify stellar companions; Oh et al. (2017) presented a model that calculates a likelihood ratio corresponding to the probability that two stars are comoving given their Gaia astrometric measurements, and they applied the model to the first Gaia data release (DR1) to identify high-confidence comoving pairs. The likelihood ratio corresponds to the ratio of two hypotheses: (1) that a pair of stars share the same physical (3D) velocity, and (2) that the two stars have independent 3D velocities. The model treats uncertainties associated with Gaia-measured proper-motion and parallax covariances in its assessment of whether two stars share the same 3D velocity. Thus, it is more robust than other methods that ignore covariances and assess whether two stars are comoving based on differences in their astrometric motions alone.

In this study, we used the Oh et al. (2017) probabilistic model to produce a catalog of Gaia Early Data Release 3 (EDR3) stellar companions to the current list of TESS Objects of Interest (TOIs). As the natural successor to the Kepler mission, TESS is currently carrying out high-precision time series photometric observations for bright stars across ~85% of the sky and has already discovered thousands of TOIs and dozens of confirmed planets (Ricker et al. 2015). Follow-up observations that confirm TOI planet candidates are ongoing, making the TOI catalog the source for the fastest-growing sample of newly identified planet hosts.

After assembling a list of TOI hosts with comoving stellar companions, we investigated the degree of alignment between the planetary and companion orbits. We used precise astrometric measurements from Gaia EDR3 to compute the angle between the relative 2D position and velocity vectors of the two comoving stars in the sky plane. We then modeled the observed distribution of angles to constrain the mutual inclinations between the planet and companion orbits, a previously unexplored geometric property of planetary systems.

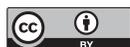

This paper is structured as follows. We provide details on the TOI catalog and our sample selection involving the Oh et al. (2017) probabilistic model in Section 2. In Section 3 we present our catalog of high-confidence TOI stellar companions obtained from applying the model to Gaia EDR3. We highlight TESS systems with stellar companions that host confirmed planets in Section 4, and we investigate possible alignment between stellar companion and planetary orbits in Section 5. We discuss these results further in Section 6.

## 2. TOI Sample and Companion Selection

We constructed our TOI host sample from the 2241 TOIs reported in Guerrero (2020) and available on the ExoFOP-TESS[3] database. The TOIs correspond to 2140 unique stellar hosts spanning TESS Sectors 1–26, whose properties are drawn from the TICv7 (Sectors 1–13) and TICv8 (Sector 14 and onward), the two most up-to-date versions of the TESS Input Catalog (TIC; Stassun et al. 2019). We searched for stellar companions to these 2140 TOI hosts within the Gaia EDR3 catalog, the latest data release resulting from the Gaia mission. Gaia EDR3 contains ∼1 million new sources compared to Gaia DR2 and features improvements in parallax and proper-motion precision of 30% and a factor of two, respectively, as well as decreased systematic uncertainties for parallaxes and proper motions by 30%–40% and a factor of ∼2.5, respectively. The photometric precision and homogeneity across color, magnitude, and celestial position are also improved (Gaia Collaboration et al. 2021).

We employed a search radius of 10′ around each TOI host to search for stellar companions. We then applied initial cuts of global parallax signal-to-noise ratio $\bar{\omega}/\sigma_{\bar{\omega}} > 4$, distance agreement $2|r_1 - r_2|/(r_1+r_2)$ to within 200%, and (point estimate) tangential velocity differences less than $|\Delta v_T| < 150$ km s$^{-1}$. These preliminary cuts are generous enough to allow for recovery of all stellar companions reported in Mugrauer (2019); their purpose is to trim down the list of potential TOI host companions without removing any true companions before applying the probabilistic model. We also note that the observational uncertainties on parameters involved in these cuts are small enough to be disregarded.

The tangential velocity calculation incorporates a point estimate of the distance derived from a correction to the Lutz–Kelker bias (Lutz & Kelker 1973):

$$\hat{r} = 1000 \left[ \frac{\bar{\omega}}{2} \left( 1 + \sqrt{1 - \frac{16}{[S/N]_{\bar{\omega}}^2}} \right) \right]^{-1} \text{ pc}, \quad (1)$$

where $\bar{\omega}$ is the parallax in mas. The tangential velocity between two stars is estimated as

$$|\Delta v_T| = |\hat{r}_1 \mu_1 - \hat{r}_2 \mu_2|, \quad (2)$$

where the proper-motion vector is $\mu = (\mu_\alpha^{*[4]}, \mu_\delta)$. We were left with ∼1,200,000 possible stellar companions following these cuts, and we plot their tangential velocity differences and separations relative to their corresponding TOI hosts in

---
[3] https://exofop.ipac.caltech.edu/tess/
[4] $\mu_\alpha^*$ is the proper-motion component in the R.A. direction, $\mu_\alpha^* = \mu_\alpha \cos\delta$.

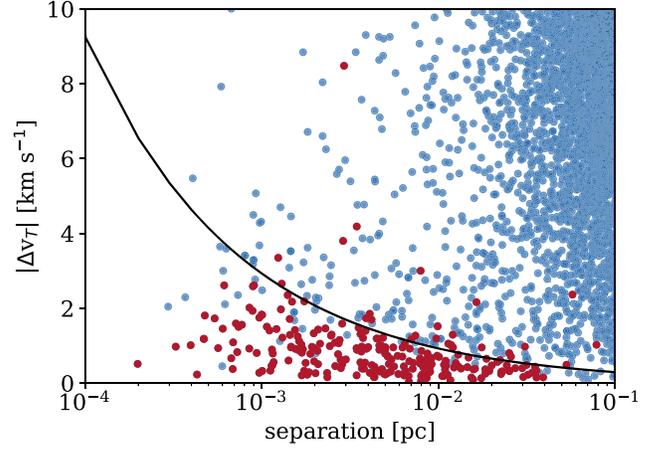

**Figure 1.** The tangential velocity differences ($\Delta|v_T|$) and 2D separations for neighboring stars relative to their TOI hosts (blue circles). The black line shows the Keplerian velocity as a function of semimajor axis for a binary pair with a total mass of 2 $M_\odot$ assuming circular orbits. The bound companions identified by our statistical analysis are shown in red.

Figure 1. We found that there is a population of stellar pairs with small separations (<1 pc) and tangential velocity differences (<2 km s$^{-1}$) that are likely gravitationally bound, also noted by Oh et al. (2017) for the Gaia DR1 sample.

We employed the probabilistic model presented in Oh et al. (2017) that incorporates reported uncertainties in the Gaia astrometric data to yield a likelihood $\mathcal{L}_1$ that a given pair of stars is comoving based on their proper motions, distances, and 3D velocities. Similarly, the likelihood that the two stars are not comoving $\mathcal{L}_2$ can be computed and compared with $\mathcal{L}_1$ to identify truly bound pairs. For more details on the model, see Oh et al. (2017). While Oh et al. (2017) established a log-likelihood ratio value of ln $(\mathcal{L}_1/\mathcal{L}_2) > 6$ as the threshold for high-confidence comoving pairs, less than half of the Mugrauer (2019) stellar companions meet this threshold, with some members of the sample yielding log-likelihood ratios as low as ln $(\mathcal{L}_1/\mathcal{L}_2) = -30$. This led us to suspect that the Oh et al. (2017) log-likelihood ratio cut was too stringent for our sample. We inspected the Mugrauer (2019) sample within the Gaia EDR3 database and found that many stellar companions have Gaia EDR3 uncertainties on astrometric data that are less than 1% of the associated proper-motion or parallax value. Such underestimated uncertainties can artificially lower the probabilistic model confidence that a pair of stars is bound. To address this, we included a "jitter" term in the proper-motion and parallax uncertainties constituting 5% of the absolute proper-motion value, and similarly 5% of the parallax value. This jitter term accounts for unknown systematic effects that may arise from factors such as the Gaia solution single-star model that fails to describe comoving binary systems. For example, TOI-837A and B have a spuriously large 3D separation of 6.6 ± 2.1 pc derived from Gaia quantities because the incorrect single-star model was assumed (Bouma et al. 2020), as further discussed in Section 4.

We ran the probabilistic model on the ∼1,200,000 possible TOI host companions and recovered ∼60,000 with ln $(\mathcal{L}_1/\mathcal{L}_2) > 0$. The purpose of this generous threshold is to retain as many true comoving systems as possible. Most of these pairs have huge projected separations on the order of $10^4$–$10^5$ au, making it unlikely that they are truly bound. To

Table 1
Properties of TOI Hosts with Stellar Companions

| TIC | R.A. (hh:mm:ss) | Decl. (deg:mm:ss) | $\mu_\alpha$ (mas yr$^{-1}$) | $\mu_\delta$ (mas yr$^{-1}$) | Parallax ($\varpi$) (mas) | $G$ (mag) |
|---|---|---|---|---|---|---|
| 332064670 | 08:52:35.22 | 28:19:47.22 | $-485.68 \pm 0.04$ | $-233.52 \pm 0.04$ | $79.45 \pm 0.04$ | 5.7 |
| 178155732 | 02:51:56.40 | $-30$:48:50.57 | $123.44 \pm 0.02$ | $106.00 \pm 0.04$ | $31.54 \pm 0.04$ | 6.3 |
| 21832928 | 17:07:55.63 | 32:06:19.07 | $-161.88 \pm 0.02$ | $-42.11 \pm 0.02$ | $27.27 \pm 0.02$ | 7.1 |
| 16740101 | 20:31:26.38 | 39:56:20.11 | $16.72 \pm 0.03$ | $20.96 \pm 0.03$ | $4.83 \pm 0.02$ | 7.6 |
| 396356111 | 00:10:23.89 | 58:29:21.98 | $-98.73 \pm 0.02$ | $-9.84 \pm 0.02$ | $10.78 \pm 0.02$ | 7.7 |
| 263003176 | 00:50:11.27 | $-83$:44:37.55 | $139.46 \pm 0.03$ | $30.39 \pm 0.02$ | $17.42 \pm 0.02$ | 7.8 |
| 243187830 | 01:07:37.99 | 22:57:10.08 | $103.17 \pm 0.04$ | $-490.28 \pm 0.02$ | $48.68 \pm 0.03$ | 8.0 |
| 371443216 | 09:50:19.22 | $-66$:06:50.14 | $5.91 \pm 0.02$ | $-15.04 \pm 0.02$ | $5.94 \pm 0.02$ | 8.2 |
| 202426247 | 15:05:49.61 | 64:02:51.71 | $-122.14 \pm 0.02$ | $110.57 \pm 0.02$ | $21.86 \pm 0.02$ | 8.2 |
| 410214986 | 23:39:39.72 | $-69$:11:45.79 | $79.53 \pm 0.01$ | $-67.55 \pm 0.02$ | $22.64 \pm 0.02$ | 8.3 |

**Note.** This table lists the TESS Input Catalog (TIC) ID and Gaia EDR3-derived R.A., decl., proper motion in the R.A. ($\mu_\alpha$) and decl. ($\mu_\delta$) directions, parallax, and $G$-band magnitudes for the 170 non-false-positive TOI hosts with detected stellar companions. The TOI hosts are sorted by their $G$-band magnitudes.

(This table is available in its entirety in machine-readable form.)

remove such pairs, we applied cuts of projected separations <10,000 au and $|\Delta v_T| < 20$ km s$^{-1}$ as motivated by the Mugrauer (2019) sample of companions, whose separations and $|\Delta v_T|$ fall within 9100 au and 20 km s$^{-1}$, respectively. We did not search separations beyond 10,000 au because such wide companions are rare and not usually predicted to remain bound throughout dynamical evolution within clusters (Parker et al. 2009; Raghavan et al. 2010). We were left with ~1500 potential companions following these cuts.

As mentioned earlier, the Gaia solution single-star model may introduce systematic errors in astrometric quantities that evade simple numerical cuts. To identify systems with such systematic errors, we visually inspected the potentially comoving systems individually, eliminating those that may still be unbound as evinced by large differences between their astrometric motions ($\Delta\mu_\alpha$ and $\Delta\mu_\delta \gtrsim$ 2–3 mas yr$^{-1}$) and distances ($\Delta r \gtrsim 50$ pc). Because our analysis does not involve a completeness study, these nonsystematic cuts should not affect our results.

Finally, we identified systems that likely harbor unresolved companions using their Gaia EDR3 renormalized unit weight error (RUWE) values. Sources with RUWE > 1.4 are generally assumed to not have well-behaved Gaia astrometric solutions assuming they are single sources (Lindegren 2018). However, because our sample is composed of systems with detected stellar companions, their RUWE values may be high because of companion contamination, making a single cut at RUWE < 1.4 inappropriate. Belokurov et al. (2020) analyzed a set of known spectroscopic binaries with RUWE values and found that RUWE rises steeply for binaries at separations <1″.5, marking this as the maximum angular separation at which companions can contaminate each other. Belokurov et al. (2020) also found that systems with tangential velocity differences exceeding their escape velocities have large RUWE, indicating an additional, closely separated companion. Thus, to identify which of our systems likely harbor unresolved companions, we made cuts on systems with separations >1″.5 that have TOI host or companion RUWE > 1.4 and tangential velocity differences greater than their escape velocities. We subsequently removed the nine systems that met these criteria from our sample. We were ultimately left with 238 high-confidence companions to 234 of the original 2140 TOI hosts, yielding a raw (without completeness corrections) detection rate of ~10.9%. These companions identified through our statistical analysis are plotted in red in Figure 1.

## 3. Stellar Companions

The majority of the 234 TOI hosts with detected companions exhibit only one companion, comprising 230 systems with two stars. There are also four triple systems. Of the 234 TOI hosts with companions, 170 have non-"FP" (false-positive), "FA" (false-alarm), or "APC" (ambiguous planetary candidate) TESS and TESS Follow-up Observing Program Working Group (TFOPWG) dispositions, whose properties are given in Table 1. The 170 TOI hosts currently marked as non-false-positive systems together harbor 172 stellar companions, whose properties are given in Table 2. Among the non-false-positive TOI systems, there are 168 binary systems and two triple systems. We noted that the TIC 37770169 triple system does not appear hierarchical based on relative separations. This system may belong to a larger group of comoving stars but failed to match with any young associations according to the Bayesian classifier BANYAN Σ (Gagné et al. 2018).

The projected separations between the TOI hosts and their companions span 40–10,000 au (our search limit), though most companions are found within 300–4000 au (Figure 2). The $G$-band magnitudes for the detected companions compared to the TOI hosts are usually fainter; the distribution of $\Delta G$ ranges from $-11.5$ to $+4.8$ and peaks at approximately $-5.0$ (Figure 3, left panel). A total of 42 of the companions to non-false-positive TOI hosts are sufficiently bright for radial velocity (RV) follow-up ($G < 13$ mag). In addition, 10 satisfy criteria required for a high level of chemical homogeneity with their TOI hosts ($\Delta T_{\rm eff} < 200$ K and $\Delta G < 0.3$ mag; Andrews et al. 2019), making them stellar twins potentially useful for studies involving differential stellar abundances. We computed TOI host and stellar companion masses for the full sample with the `isoclassify` code, which performs stellar classification with isochrone fitting by making use of astrometric and magnitude information (Huber 2017). `isoclassify` failed to predict masses for 35 of the 238 stellar companions (~15% fail rate), likely because of close companion separations that result in blending. The predicted masses for the remaining 203 stellar companions are in the range ~0.11–1.51 $M_\odot$, with ~55% exhibiting masses below 0.5 $M_\odot$. The TOI host/companion mass ratios span ~0.1–1.17 and peak at ~0.3, indicating that

Table 2
Properties of Stellar Companions to TOI Hosts

| TIC | sep (au) | $\Delta\mu_\alpha$ (mas yr$^{-1}$) | $\Delta\mu_\delta$ (mas yr$^{-1}$) | $\Delta$parallax ($\bar{\omega}$) (mas) | $\Delta G$ (mag) | $M_*$ ($M_\odot$) | ln ($\mathcal{L}_1/\mathcal{L}_2$) |
|---|---|---|---|---|---|---|---|
| 175532955 | 1828 | 0.25 ± 0.02 | 0.09 ± 0.02 | 0.05 ± 0.02 | −2.3 | $0.457^{+0.030}_{-0.030}$ | 12.96 |
| 1551345500 | 207 | 1.65 ± 0.03 | 3.67 ± 0.04 | 0.06 ± 0.02 | 1.0 | $0.604^{+0.029}_{-0.029}$ | 9.43 |
| 306362738 | 441 | 0.48 ± 0.08 | 0.72 ± 0.11 | 0.10 ± 0.10 | −5.6 | $0.296^{+0.033}_{-0.033}$ | 8.34 |
| 73649615 | 955 | 0.80 ± 0.03 | 0.30 ± 0.03 | 0.01 ± 0.03 | −1.3 | $0.265^{+0.024}_{-0.024}$ | 7.31 |
| 23434737 | 763 | 0.01 ± 0.43 | 0.51 ± 0.38 | 0.11 ± 0.48 | −11.5 | $0.112^{+0.004}_{-0.004}$ | 7.29 |
| 281781375 | 978 | 0.74 ± 0.02 | 0.18 ± 0.02 | 0.00 ± 0.02 | −1.2 | $0.782^{+0.038}_{-0.038}$ | 6.97 |
| 236445129 | 16184 | 0.47 ± 0.01 | 0.08 ± 0.02 | 0.00 ± 0.02 | −0.4 | $1.191^{+0.172}_{-0.172}$ | 6.84 |
| 452866790 | 1346 | 3.78 ± 0.07 | 0.19 ± 0.05 | 0.16 ± 0.07 | −3.5 | $0.111^{+0.004}_{-0.004}$ | 6.78 |
| 322307342 | 11838 | 0.28 ± 0.07 | 0.16 ± 0.07 | 0.06 ± 0.07 | −5.2 | ... | 6.49 |
| 280095254 | 114 | 1.01 ± 0.14 | 0.65 ± 0.21 | 0.35 ± 0.09 | −1.5 | $0.834^{+0.041}_{-0.042}$ | 6.45 |

**Note.** This table lists the TIC ID and differences in Gaia EDR3-derived R.A., decl., proper motion in the R.A. ($\mu_\alpha$) and decl. ($\mu_\delta$) directions, parallax, and G-band magnitude for the 172 stellar companions to non-false-positive TOI hosts, as well as their masses derived from isoclassify (Huber 2017). The companions are sorted by their log-likelihood ratios ln ($\mathcal{L}_1/\mathcal{L}_2$).

(This table is available in its entirety in machine-readable form.)

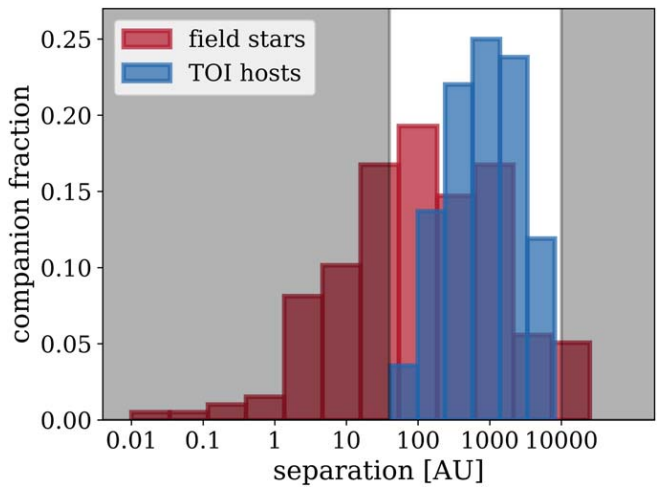

**Figure 2.** The distribution of projected separations for the 170 non-false-positive TOI hosts and their 172 detected stellar companions (blue). The red histogram corresponds to the sample of companions to solar-like field stars reported in Raghavan et al. (2010). The separations of TOI hosts and companions span 40–10,000 au (our search limit, delineated by gray shading), though most companions are found within 300–4000 au. The lack of TOI host companions at separations <40 au is due to the Gaia resolution limit of 0″.7, though the number of missed companions may be negligible (see Section 6).

the stellar companions are predominantly low-mass main-sequence stars (Figure 3, right panel).

To construct a planetless comparison sample, we drew from the TICv8 Candidate Target List (CTL). A subset of the TICv8 CTL targets are selected for TESS observations based on their ranked priorities that incorporate the probability of detecting small planet transits considering the host star radius, the total expected photometric precision, and the number of TESS sectors that may include the target. For more details on the TICv8 CTL priority calculation, see Stassun et al. (2019). We selected TICv8 CTL targets with priorities above 0.005 (yielding <200,000 stars) and $T_{\rm eff}$, mass ($M_*$), radius ($R_*$), distance ($r$), and G-band magnitude within the 1$\sigma$ bounds of the 2140 TOI system hosts. The TOI host stellar parameters were computed with isoclassify, which failed on 56 of the 2140 TOI hosts, again likely because of close companion separations that result in blending, leaving 2084 TOI hosts for this analysis. A search in the TICv8 CTL using the 1$\sigma$ bounds of the TOI systems yielded ∼19,900 targets. This sample was further trimmed down with the following similarity metric:

$$D^2 = \left(\frac{\Delta T_{\rm eff}}{\sigma_{T_{\rm eff}}}\right)^2 + \left(\frac{\Delta M_*}{\sigma_{M_*}}\right)^2 + \left(\frac{\Delta R_*}{\sigma_{R_*}}\right)^2 + \left(\frac{\Delta r}{\sigma_r}\right)^2 + \left(\frac{\Delta G}{\sigma_G}\right)^2, \quad (3)$$

which incorporates the relative $T_{\rm eff}$, $M_*$, $R_*$, $r$, and G-band magnitude between the two stars and their associated errors added in quadrature. We constructed the comparison sample by selecting the five CTL targets with the smallest comparison metric value relative to each TOI host in the 2084 TOI host sample. After removing duplicates, we were left with 4146 TICv8 CTL stars. We then recovered 703 stellar companions to 679 of the 4146 TICv8 CTL targets searched within Gaia EDR3 with the same procedure we used to identify stellar companions to TOI systems. This sample of companions to planetless CTL stars appears to be a good comparison sample for the TOI host companions by virtue of exhibiting similar distributions in parameters such as separation and brightness difference.

### 4. TOI and Confirmed Planets

The 234 TOI hosts with detected stellar companions host a total of 254 TOIs, whose properties are reported in Table 3. We plot these 254 TOIs with the total sample of 2140 TOIs in Figure 4 using the planetary orbital periods and radii.

To determine how many of our TOI hosts with stellar companions harbor confirmed planets, we selected those with TESS and TFOPWG dispositions of "CP" (confirmed planet) or "KP" (known planet) and cross-matched our catalog with the NASA Exoplanet Archive[5] (Akeson et al. 2013) and the living catalog of confirmed planets discovered by TESS.[6] As of 2022 January 9, this yielded 99 confirmed planets among the 254

---
[5] https://exoplanetarchive.ipac.caltech.edu/
[6] https://tess.mit.edu/publications//

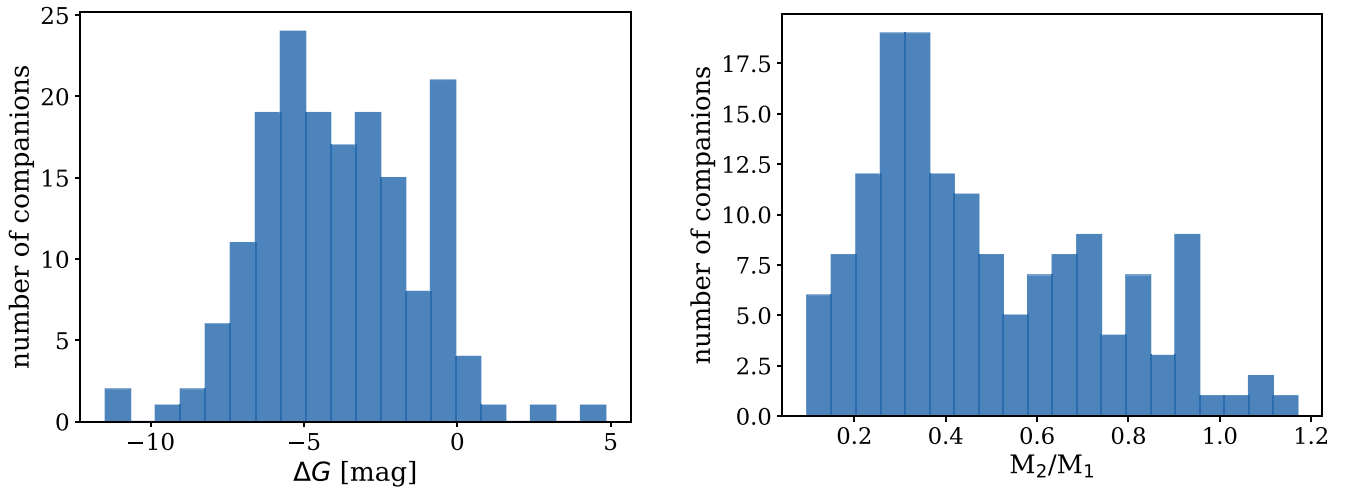

**Figure 3.** The distribution of $\Delta G$ (left) and mass ratios (right) for all 172 detected companions compared to their non-false-positive TOI hosts. The companions are comparatively fainter than the TOI hosts, with $\Delta G$ ranging from $-11.5$ to $+4.8$ and peaking at $\sim -5.0$. The mass ratio distribution of the companions relative to the TOI hosts spans $\sim 0.10$–$1.17$ and peaks at $\sim 0.3$, identifying approximately half of the companions as low-mass main-sequence stars.

TOIs, marked with an asterisk next to their corresponding TIC in Table 3. The confirmed systems span a wide range of architectures, including one planet in the Neptune desert (Díaz et al. 2020) and a young hot Jupiter system (Bouma et al. 2020). We discuss these systems in more detail below.

We detected one previously unknown stellar companion (Gaia EDR3 6520880036122448000, hereafter referred to as TOI-132B) to TOI-132, a $G = 11.3$ G dwarf that hosts a planet in the Neptune desert (Díaz et al. 2020). TOI-132B is faint ($G = 18.4$) and located at a projected separation of 3231 au (19.″6). Díaz et al. (2020) also searched for nearby companions with speckle imaging using HRCam on the 4.1 m Southern Astrophysical Research (SOAR) telescope; they noted a potential companion at 0.″079 but ultimately ruled it out based on visual inspection of the individual spectra.

We also detected one previously unknown stellar companion (Gaia EDR3 5251470948231619200, hereafter referred to as TOI-837B) at a close projected separation of 330 au (2.″3) to TOI-837, a young hot Jupiter host ($P = 0.83$ days, $R_P = 0.77^{+0.09}_{-0.07}$ $R_J$; Bouma et al. 2020). TOI-837 is a member of the $35^{+11}_{-5}$ Myr southern open cluster IC 2602, making its hot Jupiter one of the youngest transiting planets currently known. Bouma et al. (2020) identified TOI-837B with Gaia DR2 imaging and noted it as another IC 2602 member but ruled it out as a bound stellar companion based on its large 3D separation of $6.6 \pm 2.1$ pc (Bouma et al. 2020; L. G. Bouma 2020, private correspondence). However, the 3D separation calculation is quite sensitive to the Gaia distance estimate, which assumes a single-star model that likely introduced systematic errors in the parallax measurement, as discussed in Section 2. Our probabilistic analysis indicates that these two stars are bound with $\ln(\mathcal{L}_1/\mathcal{L}_2) \approx 4.09$.

### 5. Planet–Companion Orbit Alignment

We sought to constrain the relative orientations between planetary and stellar companion orbits in our sample, i.e., the mutual inclination between the two orbital axes, which we refer to as $\alpha$. Using the Gaia EDR3 equatorial coordinates (R.A. and decl.) and proper motions, we were able to precisely measure the 2D relative position and velocity vectors in the sky plane between the planet host and its comoving stellar companion.

The angle between these vectors, which we refer to as $\gamma$, encodes information on the true planet–companion mutual inclination $\alpha$. Because all planets in our sample are transiting and thus have orbital inclinations close to 90°, it follows that if the planet and companion orbits are well aligned (low $\alpha$ values), the orbital axis of the companion will also likely reside in the sky plane. In other words, if the companion follows an edge-on orbit, the angle between the 2D relative position and velocity vectors of the planet host and companion $\gamma$ will be near 0° or 180°. In contrast, if the companion and transiting planet orbits are misaligned (large $\alpha$ values), the binary orbit will more likely reside in orientations approaching face-on with $\gamma$ near 90° (Figure 5). Thus, the $\gamma$ distribution can be used to deduce the underlying distribution of $\alpha$. Figure 6 illustrates application of this method to TOI-1473. We note that the quantity $\gamma$ or "Linear Motion Parameter" was also used by Tokovinin & Kiyaeva (2016) to constrain the eccentricity distribution of wide binaries (>30 au). They modeled the eccentricity probability density distribution as $p(e) \approx 1.2e + 0.4$ with $\langle e \rangle = 0.59 \pm 0.02$, which we subsequently adopted for our investigation of mutual inclinations between planetary and companion orbits. Similarly, Hwang et al. (2021) employed the 2D relative position and velocity vectors to derive eccentricity distributions of wide binaries and found evidence for two distinct formation pathways operating at different separation regimes. The $p(e) \approx 1.2e + 0.4$ eccentricity distribution model from Tokovinin & Kiyaeva (2016) used in our alignment analysis is derived from a clean sample of 477 binaries that lack planets and have no preferential orientation. It also well represents the eccentricity distributions derived by Hwang et al. (2021) for separations of $\sim$100–3000 au, which spans the bulk of separations in our sample (Figures 2 and 7). Thus, the Tokovinin & Kiyaeva (2016) model is an appropriate prior for our entire sample of TOI systems with stellar companions.

We limited our analysis to the 168 binaries in our TOI sample, as $\gamma$ is only measurable in two-body systems. We verified that the 2D relative velocity between the two stars did not exceed their expected orbital velocity $2G(M_1 + M_2)/a$ considering their 2D projected separation and assuming circular orbits. Following this cut, we were left with 159 of the 168 binaries that together host 179 TOIs. To create a larger

Table 3
Properties of TOIs with Stellar Companions

| TIC | TOI | Num. Companions in System | TESS Dis. | TFOPWG Dis. | Source | Period (days) | $R_P$ ($R_\oplus$) | $T_{eq}$ (K) | $\ln(\mathcal{L}_1/\mathcal{L}_2)$ |
|---|---|---|---|---|---|---|---|---|---|
| 175532955 | 929.010 | 1 | PC | PC | qlp | 5.830 ± 0.000 | 2.710 ± 0.240 | 774.000 | 12.960 |
| 1551345500 | 1764.010 | 1 | PC | PC | spoc | 47.390 ± 0.000 | 13.320 ± 1.170 | 321.120 | 9.430 |
| 306362738* | 479.010 | 1 | KP | KP | spoc | 2.780 ± 0.000 | 12.680 ± 0.610 | 1270.340 | 8.340 |
| 73649615 | 756.010 | 1 | PC | PC | spoc | 1.240 ± 0.000 | 2.870 ± 0.390 | 827.940 | 7.310 |
| 23434737* | 1203.010 | 1 | CP | CP | spoc | 25.520 ± 0.004 | 2.960 ± 0.270 | 662.890 | 7.290 |
| 281781375 | 204.010 | 1 | PC | PC | spoc | 43.830 ± 0.006 | 2.510 ± 1.930 | 507.340 | 6.970 |
| 236445129* | 1282.010 | 1 | KP | KP | spoc | 0.970 ± 0.000 | 16.740 ± 0.770 | 2332.220 | 6.840 |
| 452866790* | 488.010 | 1 | CP | CP | spoc | 1.200 ± 0.000 | 1.120 ± 1.520 | 704.630 | 6.780 |
| 322307342* | 117.010 | 1 | KP | KP | spoc | 3.590 ± 0.000 | 16.470 ± 0.920 | 1658.390 | 6.490 |
| 280095254 | 235.010 | 1 | PC | PC | spoc | 10.090 ± 0.000 | ⋯ | 795.970 | 6.450 |

**Note.** This table lists the properties of all 254 TOIs with detected stellar companions. These include the TIC ID, TOI ID, number of detected Gaia EDR3 companions in the system, TESS and TFOPWG dispositions denoting candidate status (e.g., FP = false positive, APC = ambiguous planetary candidate, PC = planetary candidate, CP = confirmed planet, KP = known planet), TESS processing pipeline source (i.e., the NASA/Ames Science Processing Operations Center (SPOC) or the TESS Science Office Quick-Look Pipeline (QLP)), orbital period, radius, and equilibrium temperature. The TOIs are sorted by their detected companion log-likelihood ratios $\ln(\mathcal{L}_1/\mathcal{L}_2)$. Confirmed planet hosts are marked with an asterisk next to their corresponding TIC.

(This table is available in its entirety in machine-readable form.)

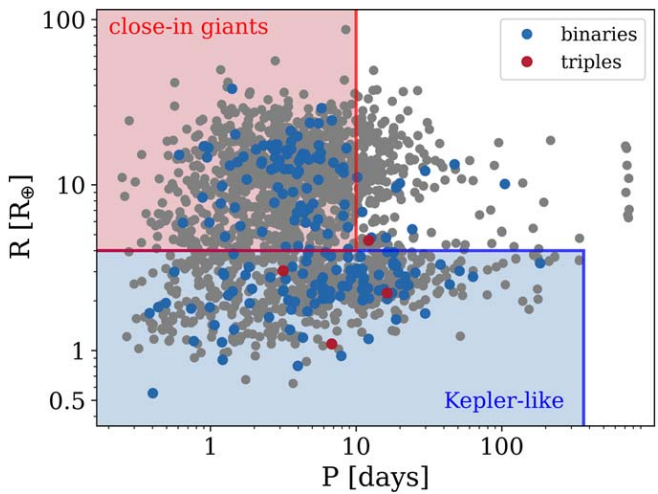

**Figure 4.** The total sample of 2140 TOIs plotted in period–radius space (gray). Non-false-positive TOIs with one detected stellar companion (binary systems) and two detected companions (triple systems) are overplotted in blue and red, respectively. There does not appear to be a preferred location for TOIs with stellar companions in period–radius space. Six TOIs fall outside this period–radius space and are not shown for easier visualization. We also delineate the regions of parameter space that harbor Kepler-like (blue) and close-in giant systems (red).

sample of planet/planet candidate-hosting binary systems, we added the Mugrauer (2019) confirmed exoplanet systems with stellar companions that pass our expected orbital velocity cuts (182 additional systems), yielding a total of 341 systems. We also constructed a comparison sample by applying our full set of cuts to the planetless CTL systems, leaving 673 systems.

For the sample of 341 planet/planet candidate systems, as well as the planetless CTL sample, we measured $\gamma$ from the equatorial coordinates and proper motions and estimated the uncertainty on $\gamma$ by sampling from multidimensional Gaussians that correspond to covariance matrices reported by Gaia EDR3. Specifically, we took 100 random draws from the covariance matrices to generate 100 iterations of $\gamma$ per system, and we took the standard deviation as $\gamma_{err}$. We then made a final cut to remove systems with $\gamma_{err} > 5°$, leaving 155 planet/planet candidate systems and 512 planetless CTL systems. We provide the $\gamma$ distribution of the planetless CTL comparison sample in the left panel of Figure 8, which appears flat. The sample of 155 binaries with $\gamma$ measurements served as the posterior sample for our hierarchical Bayesian analysis outlined in Section 5.1. We were able to ignore the influence of transiting planets on the astrometry of comoving stars because they will not generate significant perturbations to the two-body Keplerian motion considering their masses and orbital periods.

We constructed theoretical $\gamma$ distributions by generating 100,000 simulated systems with the Keplerian solver implemented in `orbitize` (Blunt et al. 2020). For each simulated system we sampled all Keplerian orbital elements uniformly in phase space and adopted the eccentricity distribution model $p(e) \approx 1.2e + 0.4$ reported by Tokovinin & Kiyaeva (2016). We assumed solar-mass stars and 1000 au separations. Supposing that the planets follow edge-on orbits ($i_p = 90°$), we can compute the orbital inclinations of the stellar companions with the mutual inclination $\alpha$. The right panel of Figure 8 displays theoretical $\gamma$ distributions generated under different assumptions. Specifically, the blue distribution corresponds to a sample of systems with circular companion orbits and no preferential alignment between the planet and companion. In this case, $\gamma$ is strongly peaked at 90°. If we allow for eccentric orbits that follow $p(e) \approx 1.2e + 0.4$, $\gamma$ is smeared out into a relatively uniform distribution between 0° and 180° as shown by the gray distribution. However, if we assume that the planet and the companion orbits are well aligned ($\alpha \lesssim 10°$), the distribution of $\gamma$ exhibits a symmetric double-peak pattern at 0° and 180° as expected from the schematic shown in Figure 5.

### 5.1. Hierarchical Bayesian Modeling

We employed a hierarchical Bayesian model (HBM) to translate the observed $\gamma$ distribution of the TOI systems to constraints on the true mutual inclination $\alpha$. We divided the TOI sample into two architecture subsamples, namely, Kepler-like systems featuring sub-Neptunes/super-Earths ($R_P \leqslant 4\ R_\oplus$ and $a < 1$ au) versus single, close-in gas giants ($P < 10$ days and $R_P > 4\ R_\oplus$) (Figure 4). These architectures were chosen based on numerous observational and theoretical lines of evidence that suggest that Kepler-like and close-in giant systems may have distinct formation pathways, e.g., giant planets strongly favor metal-rich environments (Fischer & Valenti 2005) while Kepler-like systems form readily in lower-metallicity environments (Petigura et al. 2018), and giant planets are often lonely and misaligned while Kepler-like planets frequently reside in multiplanet systems with low mutual inclinations (e.g., Winn & Fabrycky 2015 and references therein).

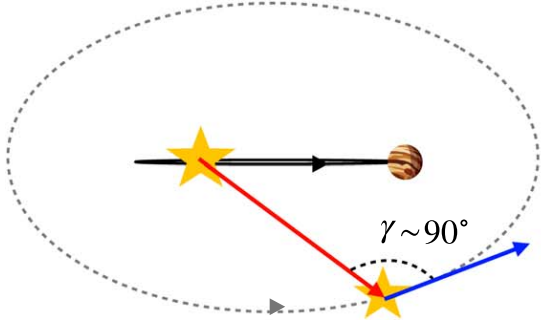
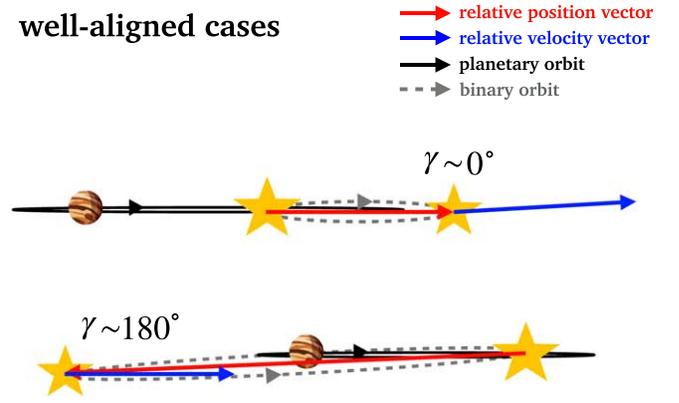

**Figure 5.** A schematic illustrating the "Linear Motion Parameter" $\gamma$, i.e., the angle between the relative 2D position and velocity vectors of the TOI and its comoving stellar companion in the sky plane. Because the transiting planet must follow a nearly edge-on orbit, $\gamma$ will favor different values depending on the mutual inclination between the orbits of the planet and companion. In the well-aligned case of small mutual inclination, the companion will likely also have an edge-on orbit, with $\gamma$ near $0°$ or $180°$. In the misaligned case of large mutual inclination, the binary orbit will tend to be face-on with $\gamma$ near $90°$.

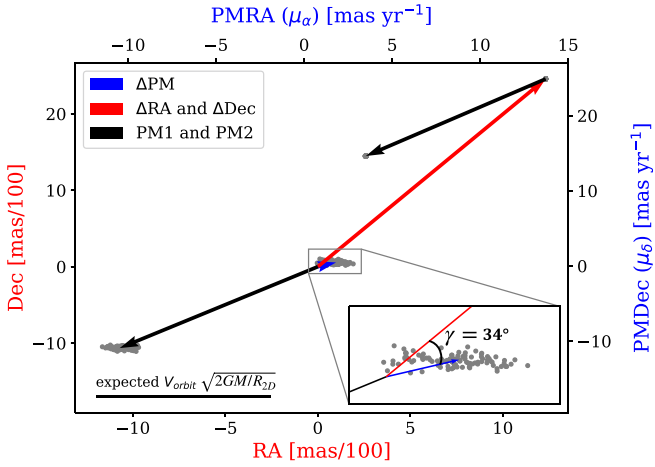

**Figure 6.** The differential position (red), differential velocity (blue), and individual velocity vectors (black) in the sky plane for one of the 155 binary TOI systems used in our alignment analysis (TOI-1473), derived from the equatorial coordinates (R.A. and decl.) and proper motions in the R.A. ($\mu_\alpha$) and decl. ($\mu_\delta$) directions. The velocity vectors are in units of mas yr$^{-1}$, while the position vector is in units of mas/100 for easier visualization. The position and velocity vector scatter is derived from sampling the covariance matrix for 100 iterations and is represented by the clouds of gray dots. The zoomed-in inset displays how $\gamma$ is measured. The expected orbital velocity (mas yr$^{-1}$) assuming circular orbits is shown for comparison and represented by the length of the black line in the lower left corner.

We employed an HBM (Hogg et al. 2010; Foreman-Mackey et al. 2014) to model the distribution of mutual inclinations between the planetary and stellar companion orbits. We considered two possible hypotheses for the $\alpha$ distributions:

1. The planetary and the stellar companion orbits are uncorrelated (no preferred $\alpha$ angle), and the resultant distribution of $\gamma$ is approximately uniform as exemplified by the gray histogram in the right panel of Figure 8.

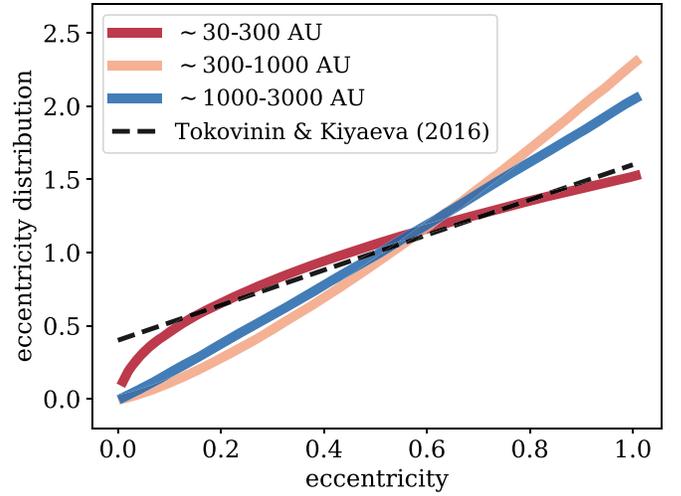

**Figure 7.** The eccentricity distribution model derived by Tokovinin & Kiyaeva (2016; black dashed) and the eccentricity distributions from Hwang et al. (2021) for binary separations in the ranges of $\sim$30–300 au (red), $\sim$300–1000 au (orange), and $\sim$1000–3000 au (blue). The Tokovinin & Kiyaeva (2016) model is in good agreement with the Hwang et al. (2021) distributions in this binary separation regime of $\sim$30–3000 au.

2. A certain fraction ($f$) of the planetary systems have a preferred orientation such that $\alpha$ can be approximated by a von Mises distribution with mean $\alpha_0$ and $\kappa$ parameter that encodes the width of the distribution, while the remaining $1 - f$ of the systems follow an isotropic $\alpha$ distribution. The resultant distribution of $\gamma$ will significantly deviate from uniformity.

We approximated the likelihood function using the $\gamma$ samples computed from the covariance matrix of each system:

$$p(\text{data}|x) \propto \prod_{k=1}^{K} \frac{1}{N} \sum_{n=1}^{N} \frac{p(\gamma_{n,k}|x)}{p_0(\gamma)}, \quad (4)$$

where data represent the observed distribution of $\gamma$, $x$ is the set of hyperparameters describing the distribution of mutual inclinations $\alpha$, $K$ is the total number of observed systems, and $N$ is the number of covariance samples. We converted the distribution of $\alpha$ described by the hyperparameters $x$ to a distribution of $\gamma$ by marginalizing over various Keplerian orbital elements. Our total set of nuisance parameters and hyperparameters includes eccentricity with the Tokovinin & Kiyaeva (2016) model taken as the prior; time of observation and argument of periapse modeled with uniform priors; stellar companion orbital inclination with a prior set by $\alpha$ and $\phi$ (an arbitrary azimuthal angle marginalized uniformly); and planetary orbital inclination, orbital period, and longitude of ascending node, all held fixed. After marginalization, we numerically evaluated $p(\gamma|x)$ and $p_0(\gamma)$ and sampled from the hyperparameter posterior distribution and Bayesian evidence $Z$ simultaneously using the nested sampling code `MultiNest` (Feroz et al. 2009). We then computed the Bayesian evidence for model comparison. We sampled the various parameters uniformly in phase space except for eccentricity, which we derived from the distribution $p(e) \approx 1.2e + 0.4$ (Tokovinin & Kiyaeva 2016).

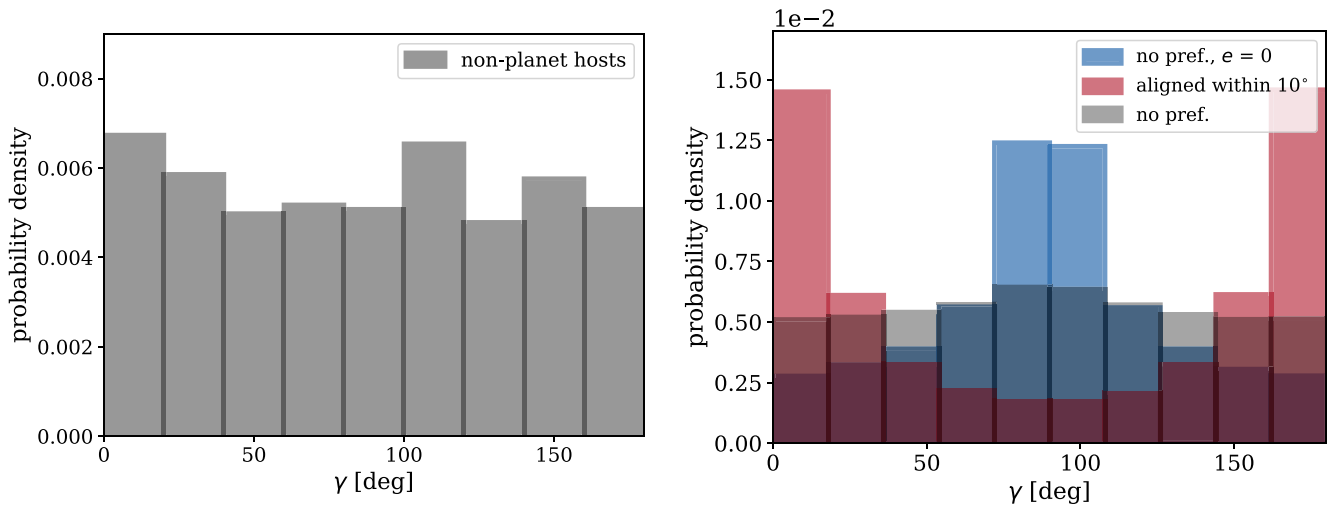

**Figure 8.** The left panel displays the normalized histogram of angles $\gamma$ between the differential position and velocity vectors in the sky plane for the 512 TICv8 CTL targets in our comparison sample that pass our $\gamma_{\rm err} < 5°$ cut (gray). The right panel displays the theoretical $\gamma$ distributions corresponding to an isotropic distribution of $\alpha$ with circular orbits (blue) or eccentric orbits (gray). The red distribution corresponds to the case of mutual inclination $\alpha$ between the transiting planet and companion agreeing within 10°. Note that the planetless comparison sample shows a rather uniform distribution of $\gamma$ as one would expect for an isotropic distribution that also accounts for orbital eccentricities.

We first tested the planetless CTL sample and found that it favors an isotropic distribution of $\gamma$ consistent with the first hypothesis. Testing the second hypothesis yielded a fraction of nonisotropic components $f$ that converged toward 0 with a 95% upper limit of $f < 18\%$, indicating that an isotropic distribution is favored again. Overall, the planetless CTL sample favors the first hypothesis with a Bayes factor of $\Delta\log(Z) = 3.1$. Considering the TOI sample, we found that Kepler-like systems slightly favor the second hypothesis with $\Delta\log(Z) = 1.1$ over the isotropic model. Specifically, for $73^{+14}_{-20}\%$ of the Kepler-like systems, the planet and companion orbits appear to favor alignment with an $\alpha$ of typically $35° \pm 24°$. Note that we combined the $\alpha_0$ and $\kappa$ posteriors together because both parameters describe the distribution of $\alpha$. The close-in giants also favor the second hypothesis with $\Delta\log(Z) = 2.0$ but, in contrast, appear to favor a perpendicular geometry, with $65^{+20}_{-35}\%$ of close-in giants exhibiting an $\alpha$ of typically $89° \pm 21°$. The best-fit $\gamma$ distributions for the Kepler-like and close-in giant systems are provided in Figure 9. These results are weakly significant according to Jeffreys (1998) and Kass & Raftery (1995), which assert that $2.5 < \log(Z) < 5$ indicates strong significance and $1 < \log(Z) < 2$ indicates positive significance. The low statistical significance of our results may be partially due to the small sizes of our Kepler-like (108 systems) and close-in giant (47 systems) samples. Moreover, $\gamma$ only provides an indirect measurement of $\alpha$. We further discuss these results and consider how the Gaia final data release may improve constraints in Section 6.

### 6. Discussion

#### 6.1. Comoving Stars for Further Characterization

We present a new catalog of Gaia EDR3 stellar companions to the 2140 unique TOI hosts from TESS Sectors 1–26. We note that the Gaia resolution limit of 0.″7 allows for companion detection if projected separations are >40 au at distances less than $\sim$60 pc. This may have contributed to lower companion completeness in our catalog; nearly all of our TOI hosts reside at farther distances, thus explaining our lack of detected stellar companions at these small separations (Figure 2). More specifically, 84% of our companions reside at separations of 300–4000 au, while only $\sim$20.5% of field star companions detected by Raghavan et al. (2010) fall in this range. However, stellar companion surveys of planet host stars yield few companions at separations interior to $\sim$100 au because planet formation is suppressed by dynamical effects from close companions (Kraus et al. 2016; Moe & Kratter 2021). This implies that the actual number of missed stellar companions to TOI hosts due to the Gaia resolution limit is likely negligible.

We found a total of 238 comoving stellar companions to 234 TOI hosts, yielding a raw companion detection rate of $\sim$10.9% with respect to the total number of 2140 searched hosts. The 234 systems include 230 binaries ($\sim$10.7%) and four triple systems ($\sim$0.19%). These fractions are lower than those of solar-type field stars, which exhibit binary and triple system fractions of $33\% \pm 2\%$ and $8\% \pm 1\%$, respectively (Raghavan et al. 2010). While our lower binary and triple detection rates likely do not stem from the Gaia resolution limit, they may be affected by Gaia pipeline incompleteness, shortcomings in our probabilistic approach, or true astrophysical differences between the companion fractions of planet host stars and field stars. The last possibility, while intriguing, cannot be entertained until the first two possibilities are ruled out.

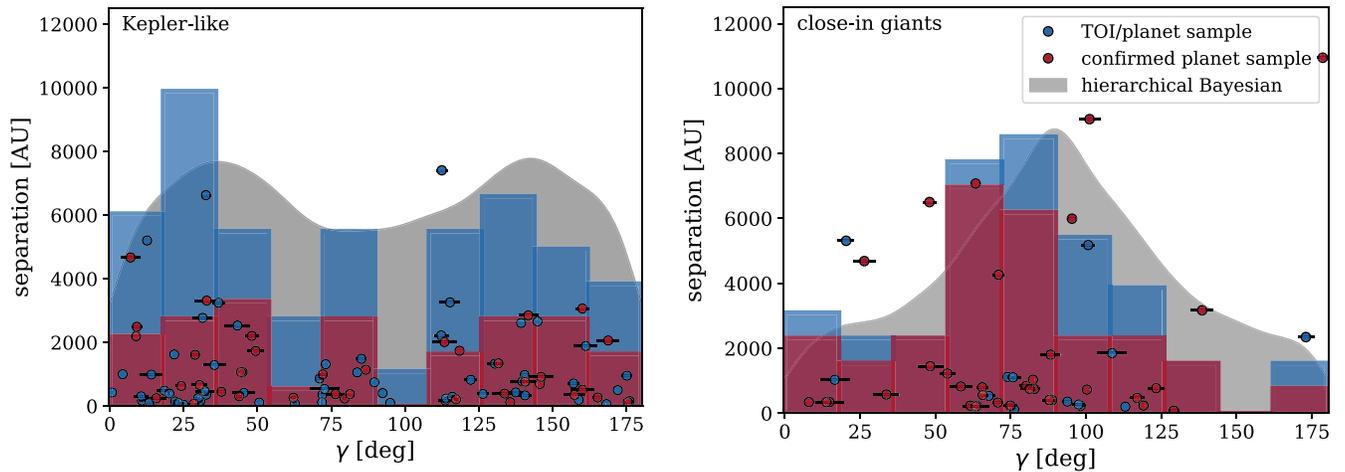

**Figure 9.** The right panel displays the scatter plot of binary separations and angles $\gamma$ between the 2D differential position and velocity vectors for the TOI (blue) and confirmed planet (red) systems that harbor close-in giant planets ($P < 10$ days and $R_P > 4\ R_\oplus$). The left panel displays the binary separations and $\gamma$ for the Kepler-like ($R_p \leqslant 4\ R_\oplus$ and $a < 1$ au) TOI and confirmed planet systems. The $\gamma$ distribution histograms are provided in the background of all plots and scaled for easier visualization. We also show the best-fit theoretical $\gamma$ distributions derived from HBM (gray).

If we consider only non-false-positive systems, there are 172 companions to 170 TOI hosts. Among these systems are 10 stellar twin binaries potentially useful for future studies involving differential stellar abundances ($\Delta T_{\rm eff} < 200$ K and $\Delta G < 0.3$ mag) and 42 systems amenable to RV follow-up for planet detection ($G < 13$ mag). Additionally, 11 of the 172 companions exhibit masses below 0.15 $M_\odot$ and distances below 100 pc, with 3 within 25 pc. This demonstrates that the probabilistic framework used in this study can identify nearby low-mass stellar companions. Such objects are valuable for surveys of stars within the solar neighborhood and their hosted planets, such as the REsearch Consortium On Nearby Stars survey within 25 pc (e.g., Henry et al. 1994) and the TRAPPIST survey of planets around nearby ultracool dwarfs (Gillon et al. 2013).

### 6.2. Thick-disk Membership

Many of the TOI hosts with stellar companions have large tangential velocities that suggest Galactic thick-disk membership. To investigate this, we computed their Galactic space motion velocities $U$, $V$, and $W$ with the procedure detailed in Johnson & Soderblom (1987) assuming the local standard of rest from Coşkunoğlu et al. (2012). Using the methodology of Reddy et al. (2006), we found that three TOI hosts with companions (TIC 166833457, TIC 175532955, TIC 23434737) have a >50% probability of belonging to the thick disk. These stars also have thick-to-thin-disk probability ratios computed from the probabilistic framework of Bensby et al. (2004, 2014) of TD/D = 4, 33, and 114, respectively (Carrillo et al. 2020). TIC 166833457 has already been confirmed as a thick-disk member via extensive chemo-kinetic follow-up (Mancini et al. 2016; Southworth et al. 2020), and its hot Jupiter candidate confirmed as WASP-98B. Though TIC 23434737 and TIC 175532955 are not confirmed to reside in the thick disk, TIC 23434737 has a TICv8 metallicity of [Fe/H] = −0.39 ± 0.05, consistent with the thick-disk population. Additionally, these stars host warm Neptune (TOI-1203) and hot super-Earth planet candidates (TOI-929), respectively, making potential thick-disk membership particularly interesting, as it would provide evidence that small, rocky planets are able to form in metal-poor environments and avoid being tidally destroyed around old stars (Buchhave et al. 2012; Hamer & Schlaufman 2020).

### 6.3. Kozai–Lidov Migration

One possible formation scenario for hot Jupiters involves a stellar companion inducing Kozai–Lidov oscillations between the hot Jupiter progenitor and its host star, leading to high-eccentricity tidal migration of the planet to its current close-in location (e.g., Fabrycky & Tremaine 2007). We considered whether the companions to our close-in giants could have instigated such Kozai–Lidov migration. Ngo et al. (2016) noted that the occurrence rate of stellar companions to hot Jupiter hosts at separations of 50–2000 au (47% ± 7%) is a factor of 2.9 higher than the rate for field stars in the same range. However, Ngo et al. (2016) also suggested that most of these companions are too far away to have instigated Kozai–Lidov migration. We performed a similar set of calculations for the close-in giants in our sample. For example, the timescale for Kozai–Lidov oscillations in the young hot Jupiter system TOI-837 is

$$\tau_{\rm KL} = \frac{2 P_b^2}{3\pi P_p^2} \frac{M_1 + M_2}{M_2} (1 - e_b^2)^{3/2} \qquad (5)$$

(Kozai 1962; Kiseleva et al. 1998), where $P_p$ is the original orbital period of the hot Jupiter progenitor around its host; $M_1$ and $M_2$ are the masses of the TOI host and stellar companion, respectively; and $P_b$ and $e_b$ are the period and eccentricity of the stellar companion, respectively. We calculated $\tau_{\rm KL}$ using `isoclassify`-predicted masses for TOI-837A and TOI-837B. We then compared $\tau_{\rm KL}$ to the timescale of general relativistic (GR) precession. This is relevant because Kozai–Lidov oscillations require a slow-changing argument of perihelion that will not occur if GR precession is sufficiently fast. We computed the expected GR precession rates for TOI-

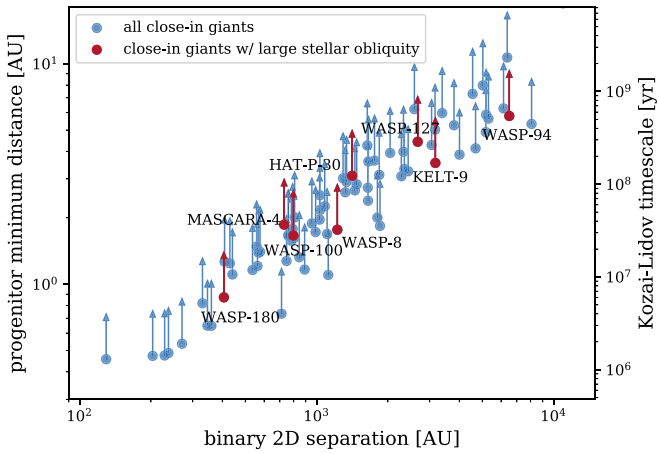

**Figure 10.** The minimum orbital distances of the close-in giant progenitors, with arrows representing the upper limit, vs. the 2D projected separation between the planet host and comoving companion in a Kozai–Lidov formation scenario. The progenitors must initially have large enough orbital periods to ensure that Kozai–Lidov oscillations induced by the comoving companion are not suppressed by GR precession. The right y-axis denotes the Kozai–Lidov timescale for the comoving binaries at their observed 2D separations. We assumed a fixed eccentricity of 0.5 for the comoving binary and fixed masses of 1 and 0.3 $M_\odot$ for the planet host and the comoving binary, respectively. These assumptions permit a one-to-one correspondence between the left and right axes. We highlight and label the planets that have both large stellar obliquities $\lambda$ and mutual inclinations $\gamma$ in red.

837 as follows:

$$\dot{\omega}_{\rm GR} = \frac{GM_*}{a_b c^2} \frac{3 n_b}{G_b^2}, \qquad (6)$$

where $n_b = 2\pi/P_b$, $G_b = \sqrt{1 - e_b^2}$, and $c$ is the speed of light. The condition for Kozai–Lidov oscillations to be suppressed by relativistic precession is $\tau_{\rm KL} \dot{\omega}_{\rm GR} > 3$ (Fabrycky & Tremaine 2007), which is well satisfied by TOI-837b for a typical choice of unknown system parameters.

However, if we allow the progenitors of close-in giant planets to form along more distant orbits, the rate of GR precession can be suppressed by several orders of magnitude. Using the 2D projected separation and `isoclassify` stellar masses, we calculated the minimum orbital distances for the progenitor planets that ensure that Kozai–Lidov oscillations are not quenched by GR precession. We found that the progenitors of close-in giants drawn from our sample of comoving binaries must have formed at minimum distances of ~0.5–10 au for Kozai–Lidov oscillations to proceed (Figure 10). Such orbital distances are consistent with the progenitors starting as cold to warm Jupiters (e.g., Dawson & Albrecht 2021; Fulton et al. 2021).

### 6.4. Planet–Companion Orbital Alignment

Stellar obliquity, or the angle between the planetary orbit and rotation axis of the host star, has been the subject of numerous exoplanet studies throughout the past decade. The diversity of the ~200 reported stellar obliquity measurements has revealed intriguing trends with respect to planet and stellar host properties (Winn et al. 2010; Mazeh et al. 2015; Louden et al. 2021) that have far-reaching implications for planet formation, migration, orbit tilting, and tidal realignment (Winn & Fabrycky 2015). In this work, we placed constraints on a similar yet distinct geometric property of planetary systems, i.e., the alignment between a planetary orbit and that of a comoving companion star. We measured the angle $\gamma$ between the 2D relative position and velocity vectors of the planet host and companion in the sky plane. This angle $\gamma$ cannot be translated to the true mutual inclination $\alpha$ on a system-by-system level owing to its dependence on other orbital elements, particularly the orbital eccentricity. However, on a population level, it is possible to marginalize over various Keplerian orbital elements and subsequently deduce the underlying $\alpha$ distribution from the observed distribution of $\gamma$.

The 2D relative velocity vector magnitude can potentially provide additional information on the true mutual inclination $\alpha$. To investigate this, we computed the normalized relative motion $\mu' = \mu/\mu^*$ for each system, where $\mu$ is the proper-motion magnitude and $\mu^*$ is the relative orbital motion (Tokovinin & Kiyaeva 2016). In Figure 11, we show the measured 2D distribution of $\mu'$ and $\gamma$. $\mu'$ is subject to more measurement uncertainty (e.g., the total mass of the binary system). Moreover, our sample size is too small to warrant 2D analysis. We chose to focus on the more informative $\gamma$ distribution.

Before assessing the observed $\gamma$ distributions, we consider a related question: do we expect any correlation between the spin axis of a star and the orbit of its comoving companion? Considering a simple core accretion model, one might assume that planets form within a protoplanetary disk that is well aligned with the spin axis of the host star. If there is a distant comoving star in the system, would the companion orbit also be coplanar with the inner planetary system? Hale (1994) argued that the rotation axis of a star and the orbit of a close-in companion within <30 au are aligned based on a comparison between $v\sin i$ measurements and stellar rotation periods. However, the situation is less clear for more distant binaries that make up the majority of our current sample. Moreover, recent analysis by Justesen & Albrecht (2020) of a larger, more well-constrained sample from TESS proved insufficient for deriving spin-binary orientations, even for close-in binaries.

If the companion orbit is uncorrelated with the planetary orbit and host star spin axis, we would expect an isotropic distribution of $\gamma$ (Figure 8, gray distribution). However, our results suggest that both Kepler-like and close-in giant systems may exhibit nonisotropic $\gamma$ distributions, though with low statistical significance ($\Delta \log(Z) = 1.1$ and 2.0, respectively). Specifically, $73^{+14}_{-20}\%$ of Kepler-like systems appear to favor alignment between the transiting planet and companion orbits, exhibiting a typical mutual inclination $\alpha$ of approximately $35° \pm 24°$. On the other hand, $65^{+20}_{-35}\%$ of close-in giants appear to favor perpendicular orientations between the transiting planet and companion, exhibiting mutual inclinations $\alpha$ that cluster around $89° \pm 21°$. As a comparison nonparametric method, we applied Kuiper's test to our architecture subsamples, which is well suited to quantifying deviations from uniformity for angular data (e.g., Fisher 1993). We derived test statistics of 0.73 and 0.58 for the close-in giant and Kepler-like systems, respectively. This indicates borderline significant deviation from uniformity for the close-in giants and less significant deviation for Kepler-like systems, in agreement with our Bayes factor results.

There are 20 systems in our sample with reported stellar obliquity $\lambda$ measurements derived using the Rossiter–McLaughlin effect (McLaughlin 1924; Rossiter 1924). Comparison of $\gamma$ and these $\lambda$ values reveals another interesting trend; the planets that are misaligned with their host stars according to obliquity measurements (nine systems with $\lambda \gtrsim 30°$) also display $\gamma$ angles near $90°$ (Figure 12). These perpendicular $\gamma$ values indicate relatively face-on orbits for the comoving binaries, which in turn betray misaligned orbits between the planets and comoving companions. We performed a simple binomial probability calculation: if we assume a uniform distribution for $\gamma$ between $0°$ and $180°$ as expected from an isotropic planet–companion orientation, the probability that all of the nine high-$\lambda$ systems will fall within the observed range of $40° < \gamma < 140°$ is about ~1%. We note that this is an a posteriori result; we devised this statistical test based on observations of the data. The true probability of finding all nine systems within the observed range is likely several times higher than 1%. Still, the correlation between $\lambda$ and $\gamma$ is obvious and suggests that close-in giants that are misaligned with their host stars are likely also misaligned with comoving companions. This result, if confirmed by future studies, may have implications for which proposed mechanisms are most effective at tilting planetary orbits (e.g., Fabrycky & Tremaine 2007; Dawson & Johnson 2018).

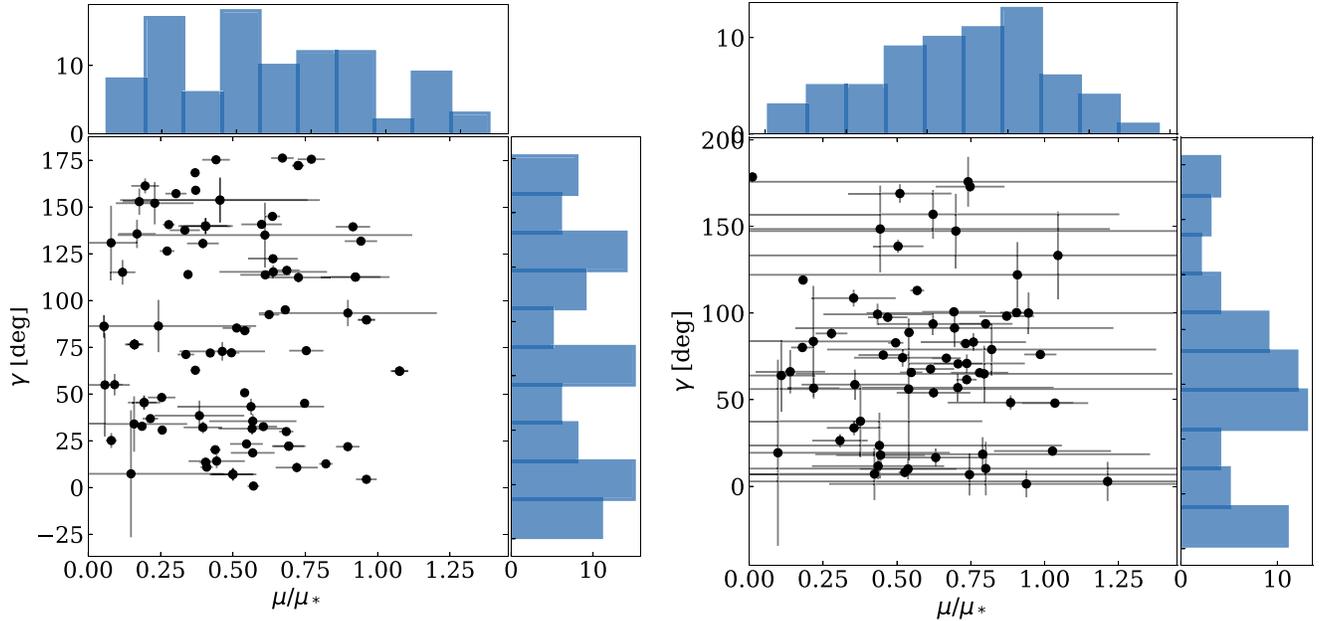

**Figure 11.** The distributions of angles $\gamma$ and normalized relative motion $\mu'$ for the TOI and confirmed planet systems that harbor close-in giant planets ($P < 10$ days and $R_P > 4\ R_\oplus$; right panel) and those with Kepler-like architectures ($R_P \leqslant 4\ R_\oplus$ and $a < 1$ au; left panel). The $\gamma$ and $\mu'$ distributions appear uncorrelated.

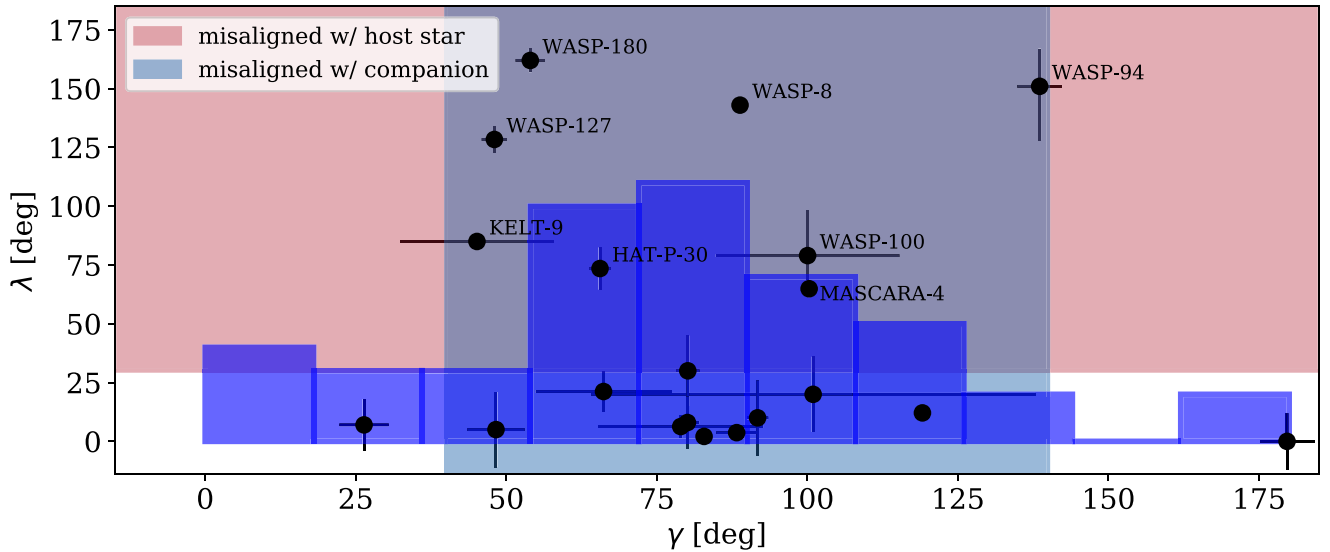

**Figure 12.** The sky-projected stellar obliquity $\lambda$ and $\gamma$ angles for the subset of 20 close-in giant planet systems in our sample with published $\lambda$ values. We overplot the histogram of all close-in giants in our sample and scale it for easier visualization (blue). Close-in giants that are misaligned with respect to their host star (red region, $\lambda > 30°$) are often also misaligned with respect to their comoving companion (blue, $40 < \gamma < 140°$).

However, we emphasize the low statistical significance of the planet–companion alignment. The Kepler-like and close-in giant systems favor nonisotropic $\gamma$ distributions with only small Bayes factor values of $\Delta \log(Z) = 1.1$ and $2.0$, respectively. This low significance can be partly attributed to the small size of our TOI sample resulting from our various data quality cuts. Moreover, Gaia EDR3 equatorial coordinates, proper motions, and parallaxes were all solved assuming a single-star model, which is bound to introduce systematic errors of varying extent into our sample of comoving binaries as discussed earlier in Section 2. Finally, $\gamma$ is an indirect proxy for the true mutual inclination $\alpha$ between the planet and companion orbits; it cannot be translated to the true planet–companion mutual inclination without knowledge of other Keplerian orbital elements.

Nevertheless, our alignment trend findings echo recent results that point to an excess of perpendicular planetary systems. In particular, Albrecht et al. (2021) found a significant preference for misaligned geometries among 57 close-in giant systems as evinced by their true, 3D obliquity measurements. If this excess of polar orbits is corroborated by future studies, it could illuminate which obliquity excitation mechanisms predominantly shape planet architectures. Future alignment analyses will also be aided by upcoming Gaia data releases; individual astrometric measurements will be provided in the full, final data release of the nominal mission that will make it possible to directly constrain orbital inclinations for a subset of our sample, thus enabling a more direct and definitive investigation of planet–companion alignment.

We thank Dan Fabrycky, Josh Winn, Simon Albrect, and Heather Knutson for insightful comments that improved the final manuscript. A.B. acknowledges funding from the National Science Foundation Graduate Research Fellowship under grant No. DGE1745301.

*Software:* Astropy (Astropy Collaboration et al. 2013, 2018).

## ORCID iDs

Aida Behmard  https://orcid.org/0000-0003-0012-9093
Fei Dai  https://orcid.org/0000-0002-8958-0683
Andrew W. Howard  https://orcid.org/0000-0001-8638-0320

## References

Akeson, R. L., Chen, X., Ciardi, D., et al. 2013, PASP, 125, 989
Albrecht, S. H., Marcussen, M. L., Winn, J. N., Dawson, R. I., & Knudstrup, E. 2021, ApJL, 916, L1
Andrews, J. J., Anguiano, B., Chanamé, J., et al. 2019, ApJ, 871, 42
Astropy Collaboration, Price-Whelan, A. M., Sipőcz, B. M., et al. 2018, AJ, 156, 123
Astropy Collaboration, Robitaille, T. P., Tollerud, E. J., et al. 2013, A&A, 558, A33
Belokurov, V., Penoyre, Z., Oh, S., et al. 2020, MNRAS, 496, 1922
Bensby, T., Feltzing, S., & Lundström, I. 2004, A&A, 415, 155
Bensby, T., Feltzing, S., & Oey, M. S. 2014, A&A, 562, A71
Blunt, S., Wang, J. J., Angelo, I., et al. 2020, AJ, 159, 89
Bouma, L. G., Hartman, J. D., Brahm, R., et al. 2020, AJ, 160, 239
Buchhave, L. A., Latham, D. W., Johansen, A., et al. 2012, Natur, 486, 375
Carrillo, A., Hawkins, K., Bowler, B. P., Cochran, W., & Vanderburg, A. 2020, MNRAS, 491, 4365
Coşkunoğlu, B., Ak, S., Bilir, S., et al. 2012, MNRAS, 419, 2844
Dawson, R. I., & Albrecht, S. H. 2021, arXiv:2108.09325
Dawson, R. I., & Johnson, J. A. 2018, ARA&A, 56, 175
Díaz, M. R., Jenkins, J. S., Gandolfi, D., et al. 2020, MNRAS, 493, 973
Dressing, C. D., Adams, E. R., Dupree, A. K., Kulesa, C., & McCarthy, D. 2014, AJ, 148, 78
Fabrycky, D., & Tremaine, S. 2007, ApJ, 669, 1298
Feroz, F., Hobson, M. P., & Bridges, M. 2009, MNRAS, 398, 1601
Fischer, D. A., & Valenti, J. 2005, ApJ, 622, 1102
Fisher, N. I. 1993, Statistical Analysis of Circular Data (Cambridge: Cambridge Univ. Press)
Foreman-Mackey, D., Hogg, D. W., & Morton, T. D. 2014, ApJ, 795, 64
Fulton, B. J., Rosenthal, L. J., Hirsch, L. A., et al. 2021, ApJS, 255, 14
Gagné, J., Mamajek, E. E., Malo, L., et al. 2018, ApJ, 856, 23
Gaia Collaboration, Brown, A. G. A., Vallenari, A., et al. 2018, A&A, 616, A1
Gaia Collaboration, Brown, A. G. A., Vallenari, A., et al. 2021, A&A, 649, A1
Gillon, M., Jehin, E., Fumel, A., Magain, P., & Queloz, D. 2013, EPJ J. Web of Conf., 47, 03001
Guerrero, N. 2020, AAS Meeting Abstracts, 235, 327.03
Hale, A. 1994, AJ, 107, 306
Hamer, J. H., & Schlaufman, K. C. 2020, AJ, 160, 138
Henry, T. J., Kirkpatrick, J. D., & Simons, D. A. 1994, AJ, 108, 1437
Hogg, D. W., Myers, A. D., & Bovy, J. 2010, ApJ, 725, 2166
Huber, D. 2017, isoclassify: v1.2, v1.2, Zenodo, doi:10.5281/zenodo.573372
Hwang, H.-C., Ting, Y.-S., & Zakamska, N. L. 2021, arXiv:2111.01789
Jeffreys, H. 1998, The Theory of Probability (Oxford: Oxford Univ. Press) https://books.google.com/books?id=vh9Act9rtzQC
Johnson, D. R. H., & Soderblom, D. R. 1987, AJ, 93, 864
Justesen, A. B., & Albrecht, S. 2020, A&A, 642, A212
Kass, R. E., & Raftery, A. E. 1995, J. Am. Stat. Assoc., 90, 773
Kiseleva, L. G., Eggleton, P. P., & Mikkola, S. 1998, MNRAS, 300, 292
Kozai, Y. 1962, AJ, 67, 591
Kraus, A. L., Ireland, M. J., Huber, D., Mann, A. W., & Dupuy, T. J. 2016, AJ, 152, 8
Lindegren, L. 2018, Gaia Technical Note GAIA-C3-TN-LU-LL-124-01, https://www.cosmos.esa.int/web/gaia/public-dpacdocuments
Louden, E. M., Winn, J. N., Petigura, E. A., et al. 2021, AJ, 161, 68
Lutz, T. E., & Kelker, D. H. 1973, PASP, 85, 573
Mancini, L., Giordano, M., Mollière, P., et al. 2016, MNRAS, 461, 1053
Matson, R., Howell, S., Horch, E., & Everett, M. 2018, AAS Meeting Abstracts, 231, 109.02
Mazeh, T., Perets, H. B., McQuillan, A., & Goldstein, E. S. 2015, ApJ, 801, 3
McLaughlin, D. B. 1924, ApJ, 60, 22
Michel, K. U., & Mugrauer, M. 2021, FrASS, 8, 14
Moe, M., & Kratter, K. M. 2021, MNRAS, 507, 3593
Mugrauer, M. 2019, MNRAS, 490, 5088
Mugrauer, M., & Michel, K. U. 2020, AN, 341, 996
Mugrauer, M., & Michel, K.-U. 2021, AN, 342, 840
Ngo, H., Knutson, H. A., Hinkley, S., et al. 2016, ApJ, 827, 8
Oh, S., Price-Whelan, A. M., Hogg, D. W., Morton, T. D., & Spergel, D. N. 2017, AJ, 153, 257
Parker, R. J., Goodwin, S. P., Kroupa, P., & Kouwenhoven, M. B. N. 2009, MNRAS, 397, 1577
Petigura, E. A., Marcy, G. W., Winn, J. N., et al. 2018, AJ, 155, 89
Raghavan, D., McAlister, H. A., Henry, T. J., et al. 2010, ApJS, 190, 1
Reddy, B. E., Lambert, D. L., & Allende Prieto, C. 2006, MNRAS, 367, 1329
Ricker, G. R., Winn, J. N., Vanderspek, R., et al. 2015, JATIS, 1, 014003
Rossiter, R. A. 1924, ApJ, 60, 15
Southworth, J., Tremblay, P.-E., Gänsicke, B. T., Evans, D., & Močnik, T. 2020, MNRAS, 497, 4416
Stassun, K. G., Oelkers, R. J., Paegert, M., et al. 2019, AJ, 158, 138
Tokovinin, A., & Kiyaeva, O. 2016, MNRAS, 456, 2070
Wang, J., Fischer, D. A., Horch, E. P., & Xie, J.-W. 2015, ApJ, 806, 248
Winn, J. N., Fabrycky, D., Albrecht, S., & Johnson, J. A. 2010, ApJL, 718, L145
Winn, J. N., & Fabrycky, D. C. 2015, ARA&A, 53, 409
Ziegler, C., Tokovinin, A., Briceño, C., et al. 2020, AJ, 159, 19
Ziegler, C., Tokovinin, A., Latiolais, M., et al. 2021, AJ, 162, 192